\documentclass[11pt]{article}

\usepackage{a4wide}
\usepackage{latexsym}
\usepackage{graphicx}
\usepackage{color}
\usepackage{hyperref}

\title{A Stochastic Model for the Evolution of the Web \\ Allowing Link Deletion}

\author{Trevor Fenner, Mark Levene, and George Loizou \\
School of Computer Science and Information Systems \\
Birkbeck College, University of London \\
London WC1E 7HX, U.K. \\ \texttt{\{trevor,mark,george\}@dcs.bbk.ac.uk}}

\date{}

\begin{document}

\maketitle

\newtheorem{theorem}{Theorem}[section]
\newtheorem{corollary}[theorem]{Corollary}
\newtheorem{lemma}[theorem]{Lemma}
\newtheorem{proposition}[theorem]{Proposition}
\newtheorem{definition}{Definition}[section]
\newtheorem{algorithm}{Algorithm}
\newtheorem{example}{Example}[section]

\begin{abstract}

Recently several authors have proposed stochastic evolutionary models for the
growth of the web graph and other networks that give rise to power-law
distributions. These models are based on the notion of preferential attachment
leading to the ``rich get richer'' phenomenon. We present a generalisation of
the basic model by allowing deletion of individual links and show that it also
gives rise to a power-law distribution. We derive the mean-field equations for
this stochastic model and show that, by examining a snapshot of the distribution
at the steady state of the model, we are able to tell whether any link deletion
has taken place and estimate the probability of deleting a link. Applying our
model to actual web graph data gives some insight into the distribution of
inlinks in the web graph and provides evidence of link deletion and the extent
to which this has occurred. Our analysis of the data also suggests a power-law
exponent of approximately $2.15$ rather than the widely published value of
$2.1$.

\end{abstract}

\section{Introduction}

Power-law distributions taking the form
\begin{equation}\label{eq:power-law}
f(i) = C \ i^{- \tau},
\end{equation}
where $C$ and $\tau$ are positive constants, are abundant in nature
\cite{SCHR91}. The constant $\tau$ is called the {\em exponent} of the
distribution. Examples of such distributions are: {\em Zipf's law}, which states
that the relative frequency of words in a text is inversely proportional to
their rank, {\em Pareto's law}, which states that the number of people whose
personal income is above a certain level follows a power-law distribution with
an exponent between 1.5 and 2, {\em Lotka's law}, which states that the number
of authors publishing a prescribed number of papers is inversely proportional to
the square of the number of publications, and {\em Gutenberg-Richter's law},
which states that the number of earthquakes over a period of time having a
certain magnitude is roughly inversely proportional to the magnitude.

\medskip

Recently several researchers have detected asymptotic power-law distributions in
the topology of the World-Wide-Web \cite{BROD00,DILL02} and, in parallel,
researchers from a variety of fields are trying the explain the emergence of
these power laws in terms of the evolution of complex networks. See
\cite{ALBE01,DORO02,NEWM03} for comprehensive surveys of recent work in this
area, detailing various stochastic models that can explain the evolution of the
web and other networks such as the internet, citation networks, collaboration
networks and biological networks. A common theme in these models is that of {\em
preferential attachment}, which results in the ``rich get richer'' phenomenon,
for example, when new links to web pages are added in proportion to the number
of currently existing links to these pages. Related approaches are the general
theoretical model covering both directed and undirected web graphs
\cite{COOP03}, the stochastic multiplicative process in which nodes appear at
different times and the rate of addition of links varies between nodes
\cite{ADAM01b}, and the statistical physics approach that uses the rate equation
technique \cite{KRAP01}.

To explain situations where pure preferential attachment models fail, we
\cite{LEVE01c} and others \cite{PENN02} have previously proposed extensions of
the stochastic model for the web's evolution in which the addition of links is
prescribed by a mixture of preferential and non-preferential mechanisms. In
\cite{LEVE01c}, we devised a general stochastic model involving the transfer of
balls between urns; this also naturally models quantities such as the numbers of
web pages in and visitors to a web site, which are not naturally described in
graph-theoretic terms. We note that our urn model is an extension of the
stochastic model proposed by Simon in his visionary paper published in 1955
\cite{SIMO55}, which was couched in terms of word frequencies in a text. We also
considered an alternative extension of Simon's model in \cite{LEVE02a} by adding
a preferential mechanism for discarding balls from urns (corresponding to
deleting web pages); this results in an exponential cutoff in the power-law
distribution.

\medskip

Our urn transfer model is a stochastic process, in which at each step with
probability $p$ a new ball (which might represent a web page) is added to the
first urn and with probability $1-p$ a ball in some urn is moved along to the
next urn. We assume that a ball in the $i$th urn has $i$ pins attached to it
(which might represent web links). It is known that the steady-state
distribution of this model is a power law, with exponent $\tau = 1 + 1/(1-p)$
\cite{SIMO55,LEVE01c}. As mentioned above, the power-law distribution breaks
down when balls may be discarded, resulting in a power-law distribution with an
exponential cutoff. So the question arises as to whether it also breaks down
when removal of pins is allowed. We answer this question here by showing that
the power-law distribution does not break down, under the constraint that more
balls are added to the system than are removed. (When the only remaining pin is
removed from a ball in the first urn that ball is removed from the system.) This
model gives a more realistic explanation for the emergence of power laws in
complex networks than the basic model without link deletion (i.e. pin removal).
Considering, for example, the web graph, the modification we make to the basic
model is that after a web page is chosen preferentially, say according to the
number of its inlinks, there is a small probability that some link to this page
will be deleted. (This is equivalent to deleting a link chosen uniformly at
random.) A possible reason for this may be that popular web pages compete for
inlinks, so that the number of inlinks a page has acquired will fluctuate with
its popularity. This is evident, for example, in the rise and fall in the
popularity of several search engines. Link deletion has also been considered by
Dorogovtsev and Mendes \cite{DORO01b} in the context of a different model in
which, at each time step, a new node is added with just one link, which
preferentially attaches itself to an existing node. In addition, at each time
step $m$ links between existing nodes are added to preferentially chosen nodes
and $c$ links between existing nodes are deleted, again preferentially. Their
conclusion that link deletion increases the power-law exponent is also obtained
here for our stochastic model.

\medskip

Consider a power law such as Lotka's law \cite{NICH89}. If this holds, then a
plot on a log-log scale of the number of authors (the {\em frequency}) against
the number of publications (the {\em value}) should reveal a straight line with
a negative slope of around $-2$. There is an obvious problem with a log-log
transformation if any of the frequencies are zero; that is to say, when there
are values $v_1$ and $v_2$, with $v_1 < v_2$, for which {\em no} author
published $v_1$ papers but at least one author published $v_2$ papers (i.e.
there is a {\em gap} in the values for the number of publications).

\smallskip

In general, we expect such gaps to occur in any data set, mainly for large
values. (This is due to the stochastic variation and the fact that the
frequencies have to be integral.) This observation is consistent with the
finding that Lotka's law does not fit well for large values (i.e. authors who
published a large number of papers) and this tail region is characterised by the
presence of gaps \cite{NICH89}.

\smallskip

One way of dealing with the problem of gaps is simply to ignore all
value-frequency pairs where the frequency is zero. However, ignoring gaps in
this way seems questionable, since the zero frequency values do give relevant
information about the data set and should not be treated as missing values. Our
unease with this approach is reinforced by the fact that computing the exponent
in this way results in a much lower value; for example, for the web inlink data
\cite{BROD00}, it gives a value around $1.5$, rather than the generally accepted
figure of $2.1$.

\smallskip

An alternative approach is to squash the non-zero frequencies up towards the
y-axis (i.e. the frequency axis), after ignoring the zero frequency values. This
is equivalent to omitting from the data set those values for which there are no
authors having that number of publications and then ranking the remaining values
in increasing order, i.e. renumbering the values starting from one. This method
also seems somewhat dubious, since the power-law relationship should really
involve the actual values rather than their ranks.

\medskip

The standard technique for fitting a power-law distribution uses linear
regression on log-log transformed data, which is not possible if gaps are
present. Another approach for handling gaps is to consider only the values up
until the first gap; this approach is only reasonable if the first gap does not
occur at too low a value. We call this the {\em unranked} approach, and note
that this approach ignores the large values in the tail of the distribution. The
unranked approach suggests one possible solution, but other approaches, such as
smoothing or using the {\em Hill plot} \cite{DREE00}, are possible. Ad-hoc
approaches have the disadvantage that they are hard to justify, and the Hill
plot, as originally defined, is only applicable after the data has been sorted,
and this also seems difficult to justify.

\smallskip

Another solution for handling gaps is the second method described above of
ranking the values with non-zero frequencies and squashing up; we call this the
{\em ranked} approach. In general, we should not expect the ranked and unranked
approaches to yield the same power-law exponent. We would expect the exponent
computed by linear regression on the log-log transformed data from the unranked
approach to be somewhat higher. One of the findings of this paper is that for
inlinks in the web graph the ranked and unranked approaches lead to a small but
noticeable difference between the exponents. (Our previous comments indicate why
we have misgivings with the ranked approach.)

\medskip

The rest of the paper is organised as follows. In Section~\ref{sec:urn} we
present an urn transfer model that extends Simon's model by allowing a pin,
chosen uniformly at random, to be discarded with some specified probability. In
Section~\ref{sec:power-law} we derive the steady state distribution of the
model, which, as stated, follows an asymptotic power law like
(\ref{eq:power-law}). In Section~\ref{sec:link} we show that, by examining a
snapshot of the distribution of balls in urns at the steady state, we are able
to tell whether removal of pins has taken place and estimate the pin removal
probability. In Section~\ref{sec:web-model} we utilise our urn model to describe
a discrete stochastic process that simulates the evolution of the degree
distribution of inlinks in the web graph. As a proof of concept, we analyse the
May 1999 data set for inlinks presented in \cite{BROD00}, which we obtained from
Ravi Kumar at IBM. We are able to show that our model is consistent with the
data and determine the extent to which link deletion has occurred. We also
investigate the discrepancy between the ranked and unranked approaches. With the
ranked approach, using linear regression on the log-log transformed data, an
exponent of $2.1$ is obtained, as in \cite{BROD00}. However, the unranked
approach results in an exponent of $2.15$, which is shown to be consistent with
our stochastic model. Although the difference between $2.1$ and $2.15$ may not
seem significant, it has been remarked in \cite{BROD00} that ``2.1 is in
remarkable agreement with earlier studies'' and in \cite{DILL02} that the
exponent is ``reliably around 2.1 (with little variation)'', which justifies us
in making a distinction between the exponents obtained using the ranked and
unranked approaches. Finally, in Section~\ref{sec:concluding} we give our
concluding remarks.

\section{An Urn Transfer Model}
\label{sec:urn}

We now present an {\em urn transfer model} for a stochastic process with urns
containing balls (which might represent web pages) that have pins (which might
represent either inlinks or outlinks) attached to them. Our model allows for
pins to be discarded with a small probability. This model can be viewed as an
extension of Simon's model \cite{SIMO55}. We note that there is a correspondence
between the Barab\'asi and Albert model \cite{BARA99b}, defined in terms of
nodes and links, and Simon's model, defined in terms of balls and pins, as was
established in \cite{BORN01}. Essentially, the correspondence is obtained by
noting that the balls in an urn can be viewed as an equivalence class of nodes
all having the same indegree (or outdegree).

\medskip

We assume a countable number of urns, $urn_1, urn_2, urn_3, \ldots \ $, where
each ball in $urn_i$ is assumed to have $i$ pins attached to it. Initially all
the urns are empty except $urn_1$, which has one ball in it. Let $F_i(k)$ be the
number of balls in $urn_i$ at stage $k$ of the stochastic process, so $F_1(1) =
1$. Then, for $k \ge 1$, at stage $k+1$ of the stochastic process one of two
things may occur:

\renewcommand{\labelenumi}{(\roman{enumi})}
\begin{enumerate}
\item with probability $p$, $0 < p < 1$, a new ball is inserted into
$urn_1$, or

\item with probability $1 - p$ an urn is selected, with $urn_i$ being
selected with probability proportional to $i F_i(k)$, and a ball is
chosen from $urn_i$; then,

\begin{enumerate}
\item with probability $r$, $0 < r \le 1$, the chosen ball is transferred
to $urn_{i+1}$, (this is equivalent to attaching an additional pin to the
ball chosen from $urn_i$), or

\item with probability $1 - r$, the ball is transferred to $urn_{i-1}$ if $i >
1$, otherwise, if $i = 1$, the ball is discarded (this is equivalent to removing
and discarding a pin from the ball chosen from $urn_i$).
\end{enumerate}
\end{enumerate}
\noindent
In the special case when $r = 1$, the process reduces to Simon's
original model.

\medskip

We note that choosing a ball preferentially (in proportion to the number of
pins) is equivalent to selecting a pin uniformly at random and choosing the ball
it is attached to. Thus, with probability $(1-p) (1-r)$, a pin chosen uniformly
at random is discarded.

\medskip

Since $i F_i(k)$ is the total number of pins attached to balls in $urn_i$, the
expected total number of pins in the urns at stage $k$ is given by
\begin{eqnarray}\label{eq:pins1}
E \Big( \sum_{i=1}^k i F_i(k) \Big) & = & 1 + (k-1) \left( p + (1-p) r -
(1-p) (1-r)
\nonumber \right) \\
& = & 1 + (k-1) \left( 1 -  2 (1-p) (1-r) \right).
\end{eqnarray}
\medskip

Correspondingly, the expected total number of balls in the urns is given by
\begin{eqnarray}\label{eq:pins2}
E \Big( \sum_{i=1}^k F_i(k) \Big) & = &  1 + (k-1) p - (1-p) (1-r)
\sum_{j=1}^{k-1} \phi_j,
\end{eqnarray}
where $\phi_j$, $1 \le j \le k-1$, is the expected probability of choosing
$urn_1$ at step (ii) of stage $j+1$, i.e.
\begin{equation}\label{eq:phi-define}
\phi_j = E \left( \frac{F_1(j)}{\sum_{i=1}^j i F_i(j)} \right).
\end{equation}
\smallskip

Now let
\begin{equation}\label{eq:phi}
\phi^{(k)} = \frac{1}{k} \sum_{j=1}^k \phi_j.
\end{equation}
\smallskip

In order to ensure that there are at least as many pins in the system as there
are balls and that, on average, more balls are added to the system than are
removed, we require the following constraint, derived from (\ref{eq:pins1}) and
(\ref{eq:pins2}),
\begin{equation}\label{eq:phi-constraint}
\frac{1 - 2 r}{1-r} \le \phi^{(k)} \le \frac{p}{(1-p) (1-r)}.
\end{equation}
\smallskip

This implies
\begin{equation}\label{eq:pr-constraint}
2 (1-p) (1-r) \le 1,
\end{equation}
which obviously holds for $r \ge 1/2$. Inequality~(\ref{eq:pr-constraint})
expresses the fact that the expected number of pins should not be negative, and
follows from (\ref{eq:pins1}).

\smallskip

We note that we could modify the initial conditions so that, for example,
$urn_1$ initially contained $\delta$ balls, $\delta > 1$, instead of just one
ball. It can be shown, from the development of the model below, that any change
in the initial conditions will have no effect on the asymptotic distribution of
the balls in the urns as $k$ tends to infinity, provided the process does not
terminate with all of the urns empty. More specifically, the probability that
the process will not terminate with all the urns empty is given by
\begin{equation}\label{eq:gambler}
1 - \left( \frac{(1-p) (1-r)}{1 - (1-p) (1-r)} \right)^\delta.
\end{equation}
\smallskip

This is exactly the probability that the gambler's fortune will increase forever
\cite{ROSS83}. (We note that (\ref{eq:gambler}) only makes sense if
(\ref{eq:pr-constraint}) holds.) Since, in practice, $r$ will be quite close to
one, once the process has survived for a few steps the probability that it will
subsequently terminate is small. From now on we will therefore assume that the
process does not terminate.

\section{Derivation of the Steady State Distribution}
\label{sec:power-law}

Following Simon \cite{SIMO55}, we now state the mean-field equations for the urn
transfer model. For $i > 1$ the expected number of balls in $urn_i$ is given by
\begin{equation}\label{eq:ss0}
E_k(F_i(k+1)) = F_i(k) + \beta_k \Big( r (i-1) F_{i-1}(k) + (1-r) (i+1)
F_{i+1}(k) - i F_i(k) \Big),
\end{equation}
where $E_k(F_i(k+1))$ is the expected value of $F_i(k+1)$, given the state of
the model at stage $k$, and
\begin{equation}\label{eq:beta-k}
\beta_k = \frac{1 - p}{\sum_{i=1}^k \ i F_i(k)}
\end{equation}
is the required normalising factor.

\medskip

In the boundary case, when $i = 1$, we have
\begin{equation}\label{eq:initial}
E_k(F_1(k+1)) = F_1(k) + p + \beta_k \Big( (1-r) \ 2 F_2(k) - F_1(k)
\Big),
\end{equation}
\smallskip
for the expected number of balls in $urn_1$, given the state at stage $k$.

\medskip

In order to obtain a solution for the model, we assume that, for large $k$, the
random variable $\beta_k$ can be approximated by a constant value
$\hat{\beta}_k$ depending only on $k$. This is defined by
\begin{displaymath}
\hat{\beta}_k = \frac{1-p}{k \left( 1 - 2 (1-p) (1-r) \right)}.
\end{displaymath}
\smallskip

The motivation for this approximation is that the denominator in the definition
of $\beta_k$, the total number of pins, has been replaced by its expectation
given in (\ref{eq:pins1}). This is a reasonable assumption since the number of
pins is the difference between two binomial random variables, and with high
probability this will be close to its expected value. We observe that using
$\hat{\beta}_k$ as the normalising factor instead of $\beta_k$ results in an
approximation similar to that of the ``$p_k$ model'' in \cite{LEVE01c}.

\medskip

Replacing $\beta_k$ by $\hat{\beta}_k$ and taking the expectations of
(\ref{eq:ss0}) and (\ref{eq:initial}), we obtain
\begin{equation}\label{eq:expected-ss0}
E(F_i(k+1)) = E(F_i(k)) + \hat{\beta}_k \Big( r (i-1) E(F_{i-1}(k)) +
(1-r)  (i+1) E(F_{i+1}(k)) - i E(F_i(k)) \Big)
\end{equation}
and
\begin{equation}\label{eq:expected-initial}
E(F_1(k+1)) = E(F_1(k)) + p + \hat{\beta}_k \Big( (1-r) \ 2 E(F_2(k)) -
E(F_1(k)) \Big),
\end{equation}
respectively.

\medskip

In order to solve (\ref{eq:expected-ss0}) and (\ref{eq:expected-initial}), we
would like to show that $E(F_i (k)) / k$ tends to a limit $f_i$ as $k$ tends to
infinity. Suppose for the moment that this is the case, then, provided the
convergence is fast enough, $E(F_i(k+1)) - E(F_i(k))$ tends to $f_i$. By ``fast
enough'' we mean that $\epsilon_{i,k+1} - \epsilon_{i,k}$ is $o(1/k)$ for large
$k$, where
\begin{displaymath}
E(F_i(k)) = k (f_i + \epsilon_{i,k}).
\end{displaymath}
\medskip

Now, letting
\begin{equation}\label{eq:beta-define}
\beta = k \hat{\beta_k} = \frac{1-p}{1 - 2 (1-p) (1-r)},
\end{equation}
we see that $\hat{\beta_k} E(F_i(k))$ tends to $\beta f_i$ as $k$ tends to
infinity. Thus, letting $k$ tend to infinity in (\ref{eq:expected-ss0}) and
(\ref{eq:expected-initial}), these yield
\begin{equation}\label{eq:ss2}
f_i = \beta \Big( r (i-1) f_{i-1} + (1-r) (i+1) f_{i+1} - i f_i \Big),
\end{equation}
and
\begin{equation}\label{eq:ss1}
f_1 = p + \beta \Big( (1-r) 2 f_2 - f_1 \Big),
\end{equation}
respectively.

\medskip

We now investigate the asymptotic behaviour of $f_i$ for large $i$. If we define
$g_i$ to be $i f_i$, we can rewrite (\ref{eq:ss2}) and (\ref{eq:ss1}) as
\begin{equation}\label{eq:ss3}
\frac{r g_{i-1}}{g_i} + \frac{(1-r) g_{i+1}}{g_i} = 1 + \frac{1}{i \beta}
\end{equation}
and
\begin{equation}\label{eq:ss3-g1}
\frac{p}{\beta g_1} + \frac{(1-r) g_2}{g_1} = 1 + \frac{1}{\beta}.
\end{equation}
\smallskip

Suppose we can write $g_i$ as
\begin{equation}\label{eq:gi}
g_i = \frac{C \ \Gamma(i+ \alpha +1)}{\Gamma(i+ \alpha + 1+\rho)} \left( 1 +
\frac{\eta}{i^2} + O \left( \frac{1}{i^3} \right) \right),
\end{equation}
where $\alpha$, $\eta$, $\rho$ and $C$ are constants.

\smallskip

Using the binomial theorem and the fact that $\Gamma(x+1) = x \Gamma(x)$, we get
\begin{eqnarray}\label{eq:ss4}
\frac{g_{i-1}}{g_i} & \approx & \frac{i+\alpha+\rho}{i+\alpha} \left( 1 +
\frac{\eta}{i^2} + O \left( \frac{1}{i^3} \right) \right) \left( 1 -
\frac{\eta}{i^2} + O \left( \frac{1}{i^3} \right) \right) \nonumber \\
& = & \frac{i+\alpha+\rho}{i+\alpha} \left( 1 + O \left( \frac{1}{i^3} \right)
\right)
\nonumber \\
& = & 1 + \frac{\rho}{i} - \frac{\rho \alpha}{i^2} + O \left( \frac{1}{i^3}
\right).
\end{eqnarray}
\medskip

Similarly,
\begin{eqnarray}\label{eq:ss5}
\frac{g_{i+1}}{g_i} & \approx & \frac{i+\alpha+1}{i+\alpha+\rho+1} \left( 1  +
O \left( \frac{1}{i^3} \right) \right) \nonumber \\
& = & 1 - \frac{\rho}{i} + \frac{\rho(\alpha+\rho+1)}{i^2} + O \left(
\frac{1}{i^3} \right).
\end{eqnarray}
\medskip

Substituting (\ref{eq:ss4}) and (\ref{eq:ss5}) into the left-hand side of
(\ref{eq:ss3}), we obtain
\begin{displaymath}
r \left( 1 + \frac{\rho}{i} - \frac{\rho \alpha}{i^2} + O \left( \frac{1}{i^3}
\right) \right) + (1-r) \left(  1 - \frac{\rho}{i} +
\frac{\rho(\alpha+\rho+1)}{i^2} + O \left( \frac{1}{i^3} \right) \right) = 1 +
\frac{1}{i \beta},
\end{displaymath}
which simplifies to
\begin{equation}\label{eq:ss7}
1 + \frac{\rho}{i} (2r -1) + \frac{\rho}{i^2} \Big( (1-r) (\rho+1) - (2r
-1)\alpha \Big) + O \left( \frac{1}{i^3} \right) = 1 + \frac{1}{i \beta}.
\end{equation}
\smallskip

For this to hold for all large enough values of $i$, we require
\begin{equation}\label{eq:ss9-rho}
\rho = \frac{1}{\beta (2 r - 1)},
\end{equation}
and
\begin{equation}\label{eq:ss9}
\alpha = \frac{(1-r) (\rho + 1)}{2 r -1}.
\end{equation}
\medskip

It is straightforward to obtain more terms in the expansion of (\ref{eq:gi}) by
a more detailed analysis, for example
\begin{displaymath}
\eta = \frac{\rho \left( (1-r) (\alpha + \rho + 1)^2 - r \alpha^2 \right)}{2 (2
r -1)}.
\end{displaymath}
\medskip

So,  with $\rho$ and $\alpha$ as in (\ref{eq:ss9-rho}) and (\ref{eq:ss9}), for
large $i$, (\ref{eq:gi}) gives an approximate solution to the recurrence defined
by (\ref{eq:ss2}) and (\ref{eq:ss1}). Thus,
\begin{equation}\label{eq:ss10}
f_i \sim C \ i^{- (1 + \rho)},
\end{equation}
where $C$ is independent of $i$ and $\sim$ means {\em is asymptotic to}.

\smallskip

From (\ref{eq:ss9-rho}) and (\ref{eq:beta-define}),
\begin{equation}\label{eq:rho}
\rho = 1 + \left( \frac{p}{1-p} \right) \left( \frac{1}{2 r - 1} \right).
\end{equation}
\smallskip


If, as is usually the case, we assume that $\rho$ is positive, then $r > 1/2$,
and it follows that increasing $p$ increases $\rho$, whereas increasing $r$
decreases $\rho$. In particular, we observe from (\ref{eq:rho}) that with pin
removal, i.e. for $r < 1$, the exponent of the power law is greater than it
would be for $r = 1$, i.e.
\begin{equation}\label{eq:rho-ge-p}
\rho \ge \frac{1}{1-p}.
\end{equation}
\smallskip

\section{Recognising Pin Removal}
\label{sec:link}

In this section we show that, using the mean-field equations derived in
Section~\ref{sec:power-law}, we are able to detect whether pin removal has taken
place by inspecting a static snapshot of the system at the steady state.

\medskip

Based on (\ref{eq:pins1}) and (\ref{eq:pins2}) we have
\begin{equation}\label{eq:solve-pins}
\frac{pins}{k} \approx 1 - 2 (1-p) (1-r),
\end{equation}
\begin{equation}\label{eq:solve-balls}
\frac{balls}{k} \approx p - (1-p) (1-r) \phi,
\end{equation}
where $pins$ and $balls$ stand for the expected numbers of pins and balls at
stage $k$, and $\phi$ is the asymptotic value of $\phi^{(k)}$, defined by
(\ref{eq:phi}), i.e. the expected proportion of pins in the first urn,
$|urn_1|/pins$. The right-hand sides of (\ref{eq:solve-pins}) and
(\ref{eq:solve-balls}) give asymptotic values of $pins$ and $balls$ as $k$ tends
to infinity. It follows that the asymptotic value of the ratio $balls/pins$,
which we denote by $\Delta$, is given by
\begin{equation}\label{eq:balls-pins}
\Delta = \frac{p - (1-p) (1-r) \phi} {1 - 2 (1-p) (1-r)}.
\end{equation}
\smallskip

From this and the fact that $\Delta \ge \phi$, it follows that
\begin{equation}\label{eq:delta1}
\phi \le \frac{p}{1 - (1-p) (1-r)} \le \Delta \le \frac{p}{1 - 2 (1-p) (1-r)}.
\end{equation}
\smallskip

Using (\ref{eq:pr-constraint}) and (\ref{eq:rho}), this implies
\begin{equation}\label{eq:delta2}
\frac{\phi}{2} \le p \le \Delta \le \frac{\rho - 1}{\rho}.
\end{equation}
\medskip

Now suppose we are given $\rho$, $\Delta$ and $\phi$. Then from (\ref{eq:rho})
and (\ref{eq:balls-pins}) we can derive the following equations:
\begin{equation}\label{eq:p-delta}
p = \frac{\phi (\rho-1)}{2 \rho (1-\Delta) - 2 + \rho \phi},
\end{equation}
\begin{equation}\label{eq:r-delta}
r = \frac{1}{2} \left( 1 + \frac{\phi}{2 \rho (1-\Delta) - 2 + \phi} \right).
\end{equation}
\medskip

Now suppose we observe an urn process with $1/2 < r \le 1$ and $0 < p < 1$, and
we take a snapshot of the system at the steady state, obtaining empirical
estimates for $\rho$, $\Delta$ and $\phi$ from the distribution of the balls in
the urns. We would like to check whether it is possible that these values could
have arisen from Simon's process, i.e. with $r=1$, with the probability of
inserting a ball into the first urn equal to, say, $p'$. So from (\ref{eq:rho})
and (\ref{eq:balls-pins}) we would require $\rho = 1/(1-p')$ and $\Delta = p'$.
Thus, from (\ref{eq:rho}), we would obtain

\begin{equation}\label{eq:simon1}
p' = \frac{\rho - 1}{\rho} = \frac{p}{1 - 2 (1-p) (1-r)},
\end{equation}
and, from (\ref{eq:balls-pins}),
\begin{equation}\label{eq:simon2}
p' = \Delta = \frac{p - (1-p) (1-r) \phi}{1 - 2 (1-p) (1-r)}.
\end{equation}
\smallskip

It is evident that (\ref{eq:simon1}) and (\ref{eq:simon2}) can only be
consistent if $r \approx 1$. Thus, simulating the urn process with $p' =
(\rho-1)/\rho$ and $r=1$ would result in the same value for $\rho$ as that
obtained from $p$ and the original value of $r$. However, in the presence of pin
removal, the value of $\Delta$ (i.e. the asymptotic value of $balls/pins$) would
be less than $p'$, the probability of inserting a new ball into the first urn.
This provides a discriminator between the two processes. As a result, by
examining the said snapshot we are able to ascertain whether the process is
consistent with the urn model in which pins may be discarded, i.e. with $r < 1$.
We apply this analysis in the next section to the degree distribution of inlinks
in the web graph.

\section{A Model for the Evolution of the Web Graph}
\label{sec:web-model}

We now describe a discrete stochastic process for simulating the evolution of
the degree distribution of inlinks in the web graph. In this model, balls
correspond to web pages and pins correspond to inlinks. At each time step the
state of the web graph is a directed graph $G = (N, E)$, where $N$ is its node
set and $E$ is its arc set. In this scenario $F_i(k)$, $i \ge 1$, is the number
of pages (nodes) in the web graph having $i$ inlinks. We note that, although we
have chosen $i$ to denote the number of inlinks, $i$ could alternatively denote
the number of outlinks, the number of pages in a web site, or any other
reasonable parameter we would like to investigate; see \cite{LEVE01c} for
further details.

\smallskip

Consider the evolution of the web graph with respect to the number of pages
having $i$ inlinks at the $k$th step of the process. Initially $G$ contains just
a single page. At each step one of three things can happen. First, with
probability $p$ a new page having one incoming link is added to $G$; this is
equivalent to placing a new ball in $urn_1$. Alternatively, with probability
$1-p$ a page is chosen, the probability of choosing a given page being
proportional to $i$, the number of inlinks the page currently has; this is
equivalent to preferentially choosing a ball from $urn_i$. Then, with
probability $r$ the chosen page receives a new inlink; this is equivalent to
transferring the chosen ball from $urn_i$ to $urn_{i+1}$. Alternatively, with
probability $1-r$ an inlink to the page is removed, and if the page has no
inlinks remaining it is removed from the graph; this is equivalent to
transferring the ball to $urn_{i-1}$ when $i > 1$, and discarding the ball from
the system if $i = 1$.

\medskip

As a proof of concept, we use the inlinks data from a large crawl of the web
performed during May 1999 \cite{BROD00}; the data set we used in this analysis
was obtained from Ravi Kumar at IBM. After removing web pages with zero inlinks
from the data set, there remain approximately 177 million pages with a total of
1.466 billion inlinks. Using the ranked approach, linear regression on the
log-log transformed data gives an exponent of $2.1052$, which is consistent with
the value of $2.1$ reported in \cite{BROD00} and confirmed subsequently in
\cite{ALBE01}; see also \cite{DILL02}.

\smallskip


As discussed in the introduction, we have some reservations about the use of the
ranked approach, as it uses ranks rather than values, thereby ignoring the gaps.
In Broder et al.'s May 1999 data set, the first gap appears at 3121, i.e. there
were no pages found with 3121 inlinks. Two possible reasons for such a gap are:
(i) there were no pages in the web having 3121 inlinks, or (ii) there were one
or more such pages but the crawl did not cover these pages. In any case, there
is no reason to believe that the web will not contain such a page in the future.
Moreover, such gaps, which are inherent in preferential attachment models,
change over time due to the growth of the web graph and the stochastic nature of
the evolutionary process.

\smallskip

\begin{figure}[ht]
\centerline{\includegraphics[width=12cm,height=9.33cm]{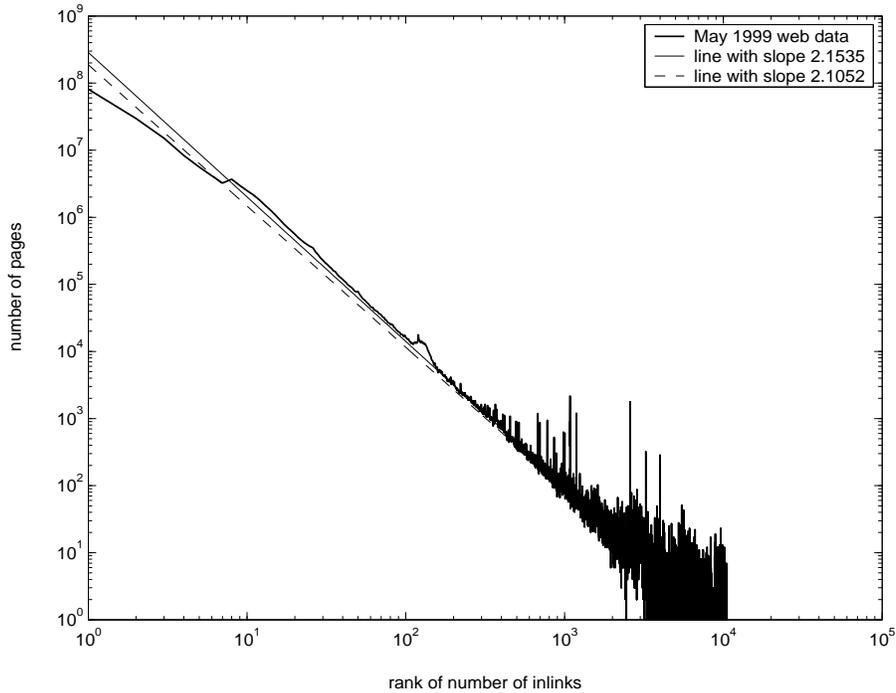}}
\caption{\label{fig:regress} Broder et al. May 1999 inlink data}
\end{figure}

One way to avoid this issue is to use the unranked approach and carry out the
regression only on the data values preceding the first gap. Using the first 3120
data values of the inlinks data set, the regression yields an exponent of
$2.1535$. In Figure~\ref{fig:regress} we show the two regression lines, which
have negative slopes of $-2.1052$, when using the ranked approach, and
$-2.1535$, when using the unranked approach. (Recall that the exponent of the
power law is $1+\rho$.)

\medskip

From the inlinks data we then estimated $\Delta$ and $\phi$, obtaining $\Delta
\approx balls/pins = 0.1205$ and $\phi \approx |urn_1|/pins = 0.0553$. Using
these estimates for $\Delta$ and $\phi$, and $\rho = 1.1535$, we computed $p$
and $r$ from (\ref{eq:p-delta}) and (\ref{eq:r-delta}), respectively, obtaining
$p = 0.0915$ and $r = 0.8280$. From the analysis in Section~\ref{sec:link}, it
is evident that link deletion is taking place in the web graph, since $r < 1$.
The extent of link deletion indicated as a proportion of insertions and
deletions is $(1-p)(1-r) = 0.156$. It would be interesting to check this value
empirically by looking at sequential snapshots of the web.

\smallskip

To validate our mean-field analysis we conducted 10 simulation runs each for $k
= 10^7$ steps with the above values of $p$ and $r$. Over the 10 runs, the mean
for $\Delta$ was $0.1197$ with standard deviation $1.1 \times 10^{-4}$, and
computing $\phi$ as $|urn_1|/pins$ gave a mean of $0.0617$ with standard
deviation $9.2 \times 10^{-5}$. However, if we compute $\phi$ from
(\ref{eq:solve-balls}) the mean is $0.0589$ with standard deviation $4.9 \times
10^{-4}$, which is closer to the empirical value of $0.0553$. We suggest that
the difference between the two estimates for $\phi$ is mainly due to the slow
convergence of $|urn_1|/pins$.



\smallskip

Lastly, we computed $\rho$ from the simulation results as
\begin{displaymath}
\frac{pins}{pins - k p},
\end{displaymath}
which follows from (\ref{eq:rho}) and (\ref{eq:solve-pins}). This gave a mean
value for $\rho$ of $1.1534$, with standard deviation $5.6 \times 10^{-5}$,
compared to the value of $1.1535$ computed from the mean-field equation
(\ref{eq:rho}).

\smallskip

For regression purposes 10 million simulation steps are insufficient to get
close to the asymptotic value of $\rho$, so we ran two additional simulations of
1 billion steps to compare the exponents obtained using the ranked and unranked
approaches. For efficiency reasons, we modified the simulation by only
considering the first $7000$ urns, approximating the effect at the upper
boundary. The simulation was implemented in Matlab, running on a desktop PC with
a 1485Mhz Intel Pentium 4 processor and 500MB of RAM, on a Windows 2000
platform. These 1 billion step simulations, with $7000$ urns, each took over 300
hours. (We conducted several more runs of 1 billion steps, varying the numbers
of urns, to validate the robustness of the approximation; a single run of 2
billion steps with $15000$ urns gave similar results to those we report below.)

For the first (second) run with $7000$ urns, we found that the first empty urn
was urn number $2403$ ($2395$) and that overall there were $5859$ ($5914$)
non-empty urns. For the unranked approach, linear regression on the log-log
transformed data for the first $2402$ ($2394$) urns gave an exponent of $2.1603$
($2.1558$). On the other hand, for the ranked approach, regression on all the
non-empty urns gave an exponent of $2.1199$ ($2.1095$). In summary, the
simulations of our stochastic model are consistent with the inlinks data set,
highlighting a small but noticeable difference in the exponent depending on
whether the ranked or unranked approach is used. In fact, with our evolutionary
model of the web graph, in common with all others, there is no easy way of
handling the gaps and, moreover, the concept of gaps has no meaning in the
context of an asymptotic mean field analysis.


\section{Concluding Remarks}
\label{sec:concluding}

We have presented an extension of Simon's classical stochastic process that
allows for pins (which might represent web links) to be discarded, and have
shown that asymptotically it still follows a power-law distribution. Given a
snapshot of a system, the mean-field equations that we have derived give
estimates of the parameters $p$ and $r$, which can then be input to our
stochastic process simulating the evolution of the system. We applied our
analysis to the May 1999 web crawl data \cite{BROD00} to detect the extent to
which link deletion had taken place. The values of $p$ and $r$ that we have
obtained indicate that approximately 15\% of all link operations are deletions.
We also ran a number of simulations to validate the mean-field analysis.

\smallskip

An interesting finding that came to light when analysing the data was that there
is good evidence that the exponent of the power law for inlinks is in fact close
to $2.15$ rather than to the widely published value of $2.1$. Although the
difference between these exponents is small, we consider it to be significant
because it suggests a more justifiable way to use linear regression to obtain
exponents from power law data.

\medskip

It would be interesting to study link deletion through historical data, such as
that provided by the {\em wayback machine} \cite{NOTE02}
(\href{http://www.waybackmachine.org}{http://www.waybackmachine.org}), in order
to gain more insight into the dynamic aspects of the evolution of the web graph.
In particular, it would be desirable to determine whether and how the exponent,
and the parameters $p$ and $r$, are changing over time.

\medskip
{\large\it Acknowledgements.} The authors would like to thank Ravi Kumar at IBM
for providing us with the inlinks data set. We would also like to thank the
referees for their constructive comments, which have helped us to improve the
presentation of the results.

\newcommand{\etalchar}[1]{$^{#1}$}

\end{document}